\documentclass[aps,pra,twocolumn,superscriptaddress,nofootinbib]{revtex4-2}
\usepackage{amsmath,amssymb,graphicx,braket,bm}
\usepackage[utf8]{inputenc}
\usepackage[colorlinks=true, linkcolor=blue, citecolor=blue, urlcolor=blue]{hyperref}
\usepackage{tikz}
\usetikzlibrary{fit,backgrounds}
\usetikzlibrary{quantikz}
\usepackage[normalem]{ulem}
\usepackage{xcolor}

\newcommand{\panelgraphic}[3]{%
    \begin{tikzpicture}
        \node[inner sep=0] (img) {%
            \includegraphics[width=#1]{#2}%
        };
        \node[
            anchor=north west,
            font=\bfseries,
            fill=white,
            inner sep=2pt
        ] at ([xshift=-2mm,yshift=-2mm]img.north west) {#3};
    \end{tikzpicture}%
}

\begin{document}

\title{Phase-Altered Interleaved Randomized Benchmarking for Compiled Non-Clifford Gates}
\author{Simona K. Grigorova}
\affiliation{Center for Quantum Technologies, Department of Physics, Sofia University, James Bourchier 5 blvd., 1164 Sofia, Bulgaria}

\author{Nikolay V. Vitanov}
\affiliation{Center for Quantum Technologies, Department of Physics, Sofia University, James Bourchier 5 blvd., 1164 Sofia, Bulgaria}

\author{Boyan T. Torosov}
\affiliation{Georgi Nadjakov Institute of Solid State Physics, Bulgarian Academy of Sciences, 72 Tsarigradsko Chaussée, 1784 Sofia, Bulgaria}

\date{\today}

\begin{abstract}
Interleaved randomized benchmarking (IRB) provides a scalable estimate
of a gate's error rate, but its standard guarantees require the
interleaved gate to be Clifford~\cite{Magesan2012Interleaved,magesan2012characterizing}.
In superconducting processors, many non-Clifford phase gates in compiled
circuits are implemented virtually as software-defined frame updates
rather than as additional control pulses~\cite{mckay2017efficient}.
This raises the question of whether inserting or removing such virtual
phases measurably changes IRB error estimates.

We introduce \emph{phase-altered interleaved randomized benchmarking}
(PA-IRB), a paired-IRB diagnostic protocol comparing phase-stripped and
phase-dressed Clifford interleaving gates derived from the same compiled
implementation. The phase-stripped construction removes the virtual
non-Clifford phase operations, whereas the phase-dressed construction
retains each such operation and immediately follows it by its inverse.
PA-IRB compares the resulting error estimates through
$\Delta r=r_{\mathrm{pd}}-r_{\mathrm{ps}}$.
The two constructions have the same ideal Clifford action, allowing their implementation-dependent differences to be examined within the standard IRB setting.

As a case study, we apply PA-IRB to an exact compiled Toffoli gate
executed on IBM Quantum \texttt{ibm\_kingston}, where the non-Clifford
$R_Z(\pm\pi/4)$ phases corresponding to $T/T^\dagger$ operations are
implemented virtually. Across 215 physical qubit triples, we obtain a mean
$\Delta r=1.65\times10^{-3}$, with fitted outcomes on both sides of zero:
169 have $\Delta r>0$ and 46 have $\Delta r<0$. These fitted differences characterize the tested population and acquisition window. More generally, PA-IRB provides a lightweight diagnostic for benchmarking workflows involving software-defined phase operations.
\end{abstract}

\maketitle

\section{Introduction}

Universal quantum computation requires the implementation of high-fidelity non-Clifford operations. While Clifford gates alone generate a group that is efficiently simulable on classical hardware, adding non-Clifford gates promotes the gate set to universality~\cite{gottesman1998heisenberg,nielsen2002quantum}. This capability is essential for realizing quantum algorithms that outperform classical counterparts. As such, it becomes crucial to develop and apply benchmarking techniques capable of characterizing non-Clifford operations.

A key feature of many quantum computing architectures is that
certain non-Clifford phase operations, most notably the $T$ gate, are implemented
not through physical control pulses but as software-defined virtual $Z$-rotations
via control-frame updates~\cite{mckay2017efficient}.
Such virtual operations can be implemented without additional gate duration and are therefore commonly treated as effectively error-free at the pulse level~\cite{mckay2017efficient}.
As a result, compiled non-Clifford gates such as the Toffoli (CCX), which decompose into physical Clifford gates supplemented by virtual phase rotations, can retain the same hardware-executed non-$R_Z$ gate skeleton while differing
in their software-defined phase structure.

These observations sit within a broader landscape of scalable benchmarking methods that increasingly
target \emph{compiled} circuit performance. For example, cycle benchmarking (CB) uses local randomization
to estimate error rates at the level of compiled layers (cycles), avoiding some overheads and failure
modes associated with large global Clifford compilations~\cite{erhard2019cyclebenchmarking}. Mirror-circuit
benchmarks provide an end-to-end characterization of compiled circuits without restricting attention to a
single interleaved operation~\cite{proctor2022mirror}. In parallel, theoretical work has clarified what RB
decay parameters quantify beyond the simplest gate-independent noise model and how systematic effects can
arise~\cite{proctor2017whatrb,helsen2022general}. Tomography-style approaches such as gate set tomography
(GST) provide detailed model-based reconstructions at higher experimental and classical cost~\cite{nielsen2021gst}.
Most closely related to the setting considered here, recent GST work has explicitly analyzed when “error-free
virtual-$Z$” (static-$VZ$) assumptions are justified on platforms where $Z$ phases are software-defined~\cite{cao2024staticvz}.

Motivated by this context, we focus on a concrete question at the software--hardware abstraction boundary:
does retaining explicitly cancelling virtual phase gates in a Clifford surrogate change the fitted IRB decay under the tested compilation and execution conditions?

In this work, we introduce \emph{phase-altered interleaved randomized
benchmarking} (PA-IRB), a paired-IRB diagnostic protocol designed to address
this question. PA-IRB compares two Clifford interleaving constructions derived
from the same compiled operation: a \emph{phase-stripped} construction
in which the virtual non-Clifford phase rotations are removed, and a
\emph{phase-dressed} construction in which those rotations are retained and
immediately followed by their inverses. The two constructions preserve the
same non-$R_Z$ physical gate skeleton, while differing in their explicit
virtual phase content.

Separate IRB experiments yield error estimates $r_{\mathrm{ps}}$ and
$r_{\mathrm{pd}}$ for the phase-stripped and phase-dressed constructions,
respectively. We define
\begin{equation}
    \Delta r = r_{\mathrm{pd}}-r_{\mathrm{ps}},
    \label{eq:delta_r}
\end{equation}
which quantifies the change in the fitted IRB error between the two ideally equivalent constructions under the employed compilation and execution conditions.

We demonstrate PA-IRB using an exact compiled Toffoli gate on the
IBM Quantum \texttt{ibm\_kingston} processor. The protocol is applied across
215 calibration-qualified physical qubit triples, providing a population-scale
comparison of the phase-stripped and phase-dressed error estimates. Across the
population, the paired estimates are concentrated near
$r_{\mathrm{pd}}=r_{\mathrm{ps}}$, while fitted outcomes occur on
both sides of this relation, including cases in which the phase-dressed
surrogate yields the lower extracted error.

The paper is organized as follows. In Sec.~II we review
randomized benchmarking and interleaved randomized benchmarking
and discuss Clifford and non-Clifford gate execution.
In Sec.~III we introduce the PA-IRB protocol.
In Sec.~IV we describe the compiled Toffoli implementation and
experimental protocol on \texttt{ibm\_kingston}.
In Sec.~V we present the experimental results, and in Sec.~VI
we discuss their implications for benchmarking compiled
quantum circuits.

The construction extends to compiled operations containing arbitrary
virtual $R_Z(\theta)$ rotations, provided that phase stripping produces
a Clifford surrogate and the intended instruction-level difference
is preserved during compilation.

\section{Background}
\subsection{Randomized Benchmarking and IRB}

Randomized benchmarking (RB) is a protocol for estimating the average error rate of a gate set by applying random sequences of gates drawn from a unitary 2-design, most commonly the Clifford group~\cite{magesan2011scalable,knill2008randomized}. Each sequence ends with an inversion gate constructed to ideally return the system to its initial state as seen in Fig.~\ref{circ1}. 

\begin{figure}[h]
  \centering
\begin{quantikz}[row sep=0.4cm, column sep=0.4cm]
  \lstick{$\ket{0}$} & \gate{C_1} & \gate{C_2} & \push{\cdots}\qw & \gate{C_m} & \gate{C_{\text{inv}}} & \meter{}
\end{quantikz}
  \caption{Randomized benchmarking circuit consisting of random Clifford gates $C_i$ of length $m$ followed by an inversion gate $C_{\mathrm{inv}}$ at the end.}
  \label{circ1}
\end{figure}

Under the assumption of gate-independent noise, the probability of measuring the initial state after a sequence of length $m$ follows an exponential decay model~\cite{magesan2011scalable,proctor2019direct}:
\begin{equation}
    P(m) = A p^m + B,
    \label{eq1}
\end{equation}
where $p$ is related to the average gate fidelity \cite{magesan2012characterizing}:
\begin{equation}
    F_{\text{avg}} = \frac{(d - 1)p + 1}{d},
\end{equation}
and \( d = 2^n \) is the Hilbert space dimension for an \( n \)-qubit system.

Interleaved randomized benchmarking (IRB) is shown in Fig.~\ref{circ2}. It extends RB by inserting a fixed gate \( G \) between Cliffords in each sequence \cite{Magesan2012Interleaved}. 

\begin{figure}[h]
  \centering
\begin{quantikz}[row sep=0.4cm, column sep=0.5cm]
  \lstick{$\ket{0}$} & \gate{C_1} & \gate{G} & \gate{C_2} & \gate{G} & \push{\cdots}\qw  \\  \push{\cdots}\qw & \gate{G} & \gate{C_m} & \gate{G} & \gate{C_{\text{inv}}} & \meter{}
\end{quantikz}
  \caption{Interleaved randomized benchmarking circuit. Again we have random Clifford gates from $C_1$ to $C_m$ but here the gate we want an error estimate for is interleaved between the random Cliffords. The circuit is followed by an inversion gate $C_{\mathrm{inv}}$.}
  \label{circ2}
\end{figure}

Under the standard IRB assumptions, the interleaved-gate error estimate is
\begin{equation}
    r_G = \frac{d - 1}{d}
    \left(1-\frac{p_G}{p}\right),
    \label{eq3}
\end{equation}
where $p$ is the reference RB decay parameter and $p_G$ is the interleaved
decay parameter~\cite{magesan2012characterizing,Magesan2012Interleaved}.
While IRB is formally justified for Clifford interleaving gates,
extensions beyond the Clifford group require additional assumptions or
structure~\cite{proctor2019direct,helsen2022general,helsen2022rbframework}.
As in standard IRB, the extracted error estimates are interpreted within the applicable noise assumptions; gate dependence and temporal variation can contribute systematic uncertainty.

\subsection{Clifford and non-Clifford gate execution}

A unitary operator \(U\) acting on \(n\) qubits is a Clifford operation if it
normalizes the \(n\)-qubit Pauli group \(\mathcal{P}_n\),
\begin{equation}
    U \mathcal{P}_n U^\dagger = \mathcal{P}_n,
\end{equation}
or equivalently, if
\begin{equation}
    U P U^\dagger \in \mathcal{P}_n
    \qquad
    \text{for all } P\in\mathcal{P}_n.
\end{equation}

For the experiments considered in this work, the three-qubit circuits are
compiled to the gate basis
\[
\left\{
    \mathrm{ID},
    R_Z(\phi),
    \sqrt{X},
    X,
    CZ
\right\}.
\]
The relevant single-qubit operations are
\begin{align}
R_Z(\phi)
    &= e^{-i\phi Z/2}
    =
    \begin{pmatrix}
        e^{-i\phi/2} & 0 \\
        0 & e^{i\phi/2}
    \end{pmatrix},
    \\[4pt]
\sqrt{X}
    &=
    e^{i\pi/4}e^{-i\pi X/4}
    =
    \frac{1}{2}
    \begin{pmatrix}
        1+i & 1-i \\
        1-i & 1+i
    \end{pmatrix},
    \\[4pt]
X
    &=
    \begin{pmatrix}
        0 & 1 \\
        1 & 0
    \end{pmatrix}.
\end{align}
Both \(X\) and \(\sqrt{X}\) are Clifford operations. In contrast,
\(R_Z(\phi)\) is Clifford only when
\begin{equation}
    \phi = k\frac{\pi}{2},
    \qquad k\in\mathbb{Z}.
\end{equation}
For a generic rotation angle,
\begin{equation}
R_Z(\phi) X R_Z(\phi)^\dagger
    =
    \cos(\phi)X+\sin(\phi)Y,
\end{equation}
which is not generally a Pauli operator and therefore places
\(R_Z(\phi)\) outside the Clifford group.
In particular, up to an irrelevant global phase,
\begin{equation}
    T \sim R_Z(\pi/4),
    \qquad
    T^\dagger \sim R_Z(-\pi/4),
\end{equation}
so the $T$ and $T^\dagger$ phase operations appearing in the compiled
Toffoli implementation are non-Clifford, whereas rotations
$R_Z(k\pi/2)$ are Clifford.

The two-qubit entangling operation used in the compiled circuits is the
controlled-\(Z\) gate,
\begin{equation}
CZ =
\begin{pmatrix}
1 & 0 & 0 & 0 \\
0 & 1 & 0 & 0 \\
0 & 0 & 1 & 0 \\
0 & 0 & 0 & -1
\end{pmatrix}.
\end{equation}
The \(CZ\) gate is Clifford, as is the identity operation.

This distinction is central to the PA-IRB construction. In the
phase-stripped surrogate, the selected non-Clifford $R_Z(\phi)$ rotations are
removed. In the phase-dressed surrogate, each is retained and immediately
followed by its inverse. The two constructions are equivalent at the level of
the ideal unitary and preserve the same non-$R_Z$ physical gate skeleton, but
the dressed circuit retains explicit virtual non-Clifford phase instructions.
PA-IRB tests whether this instruction-level difference leads to a measurable
difference in the extracted IRB error.

For the compiled circuits considered here, the basis
$\{\mathrm{ID},R_Z(\phi),\sqrt{X},X,CZ\}$ therefore contains a single
potentially non-Clifford component: $R_Z(\phi)$ for
$\phi\neq k\pi/2$.

\section{Phase-Altered Interleaved Randomized Benchmarking}

Phase-altered interleaved randomized benchmarking (PA-IRB) is a paired
diagnostic protocol designed to test whether the explicit presence of virtual
non-Clifford phase instructions modifies the error extracted by IRB.

\begin{figure}[h]
  \centering
\begin{quantikz}[row sep=0.30cm, column sep=0.40cm]
  \lstick{$\ket{0}^{\otimes n}$} &
  \gate{K_1} &
  \gate{Z_{\mathrm{Cl}}^{(1)}} &
  \gate{Z_{\mathrm{nCl}}^{(1)}} &
  \gate{K_2} &
  \gate{Z_{\mathrm{Cl}}^{(2)}} &
  \gate{Z_{\mathrm{nCl}}^{(2)}} &
  \push{\cdots}\qw & \\
  \push{\cdots}\qw &
  \gate{K_L} &
  \gate{Z_{\mathrm{Cl}}^{(L)}} &
  \gate{Z_{\mathrm{nCl}}^{(L)}} &
  \qw
\end{quantikz}
  \caption{Schematic compiled operation $G$ decomposed into Clifford blocks $K_\ell$, Clifford phase updates $Z_{\mathrm{Cl}}^{(\ell)}$ (e.g., $R_Z(k\pi/2)$), and non-Clifford phase updates $Z_{\mathrm{nCl}}^{(\ell)}$ (e.g., $R_Z(\pi/4)$, $T$, $T^\dagger$).}
  \label{circ:G_general}
\end{figure}

Starting from the same frozen compiled operation $G$, shown
schematically in Fig.~\ref{circ:G_general}, PA-IRB constructs
two Clifford interleaving circuits:
\begin{enumerate}
    \item A \emph{phase-stripped} construction $G_\mathrm{ps}$
(Fig.~\ref{circ:G_ps}), obtained by removing the selected non-Clifford
virtual phase rotations.

\begin{figure}[h]
  \centering
\begin{quantikz}[row sep=0.30cm, column sep=0.40cm]
 \lstick{$\ket{0}^{\otimes n}$} &
  \gate{K_1} &
  \gate{Z_{\mathrm{Cl}}^{(1)}} &
  \gate{K_2} &
  \gate{Z_{\mathrm{Cl}}^{(2)}} &
  \qw &
  \push{\cdots}\qw & \\
   \push{\cdots}\qw & 
  \gate{K_L} &
  \gate{Z_{\mathrm{Cl}}^{(L)}} &
  \qw
\end{quantikz}
  \caption{Phase-stripped implementation $G_{\mathrm{ps}}$ obtained from $G$ by removing all non-Clifford phase updates $Z_{\mathrm{nCl}}^{(\ell)}$.}
  \label{circ:G_ps}
\end{figure}
    
   \item A \emph{phase-dressed} construction $G_{\mathrm{pd}}$
(Fig.~\ref{circ:pd_pairs}), in which each selected non-Clifford phase
rotation is retained and immediately followed by its inverse,
\begin{equation}
    Z_{\mathrm{nCl}}^{(\ell)}
    \longrightarrow
    Z_{\mathrm{nCl}}^{(\ell)}
    Z_{\mathrm{nCl}}^{(\ell)\dagger}.
\end{equation}

    \begin{figure}[h]
  \centering
\begin{quantikz}[row sep=0.30cm, column sep=0.45cm]
    \lstick{$\ket{0}^{\otimes n}$} &
  \gate{K_1} &
  \gate{Z_{\mathrm{Cl}}^{(1)}} &
  \gate{Z_{\mathrm{nCl}}^{(1)}} &
  \gate{Z_{\mathrm{nCl}}^{(1)\dagger}} &
  \qw & \\
  \qw &
  \gate{K_2} & 
  \gate{Z_{\mathrm{Cl}}^{(2)}} &
  \gate{Z_{\mathrm{nCl}}^{(2)}} &
  \gate{Z_{\mathrm{nCl}}^{(2)\dagger}} &
  \push{\cdots}\qw & \\
  \push{\cdots}\qw &
  \gate{K_L} &
  \gate{Z_{\mathrm{Cl}}^{(L)}} &
  \gate{Z_{\mathrm{nCl}}^{(L)}} &
  \gate{Z_{\mathrm{nCl}}^{(L)\dagger}} &
  \qw
\end{quantikz}
  \caption{Phase-dressed construction where non-Clifford phase updates appear in commuting
  pairs $Z_{\mathrm{nCl}}^{(\ell)} Z_{\mathrm{nCl}}^{(\ell)\dagger} = I$ (e.g., $T T^\dagger$). Since all
  $Z$-axis phase operations shown are diagonal in the computational basis, they mutually commute
  and may be reordered within each phase layer. The dressed and stripped constructions therefore have the same
ideal unitary. PA-IRB tests whether retaining the explicit
cancelling phase instructions changes the extracted IRB error.}
  \label{circ:pd_pairs}
\end{figure}
\end{enumerate}

IRB is performed separately for $G_{\mathrm{ps}}$ and
$G_{\mathrm{pd}}$, yielding error estimates $r_{\mathrm{ps}}$
and $r_{\mathrm{pd}}$. Their difference quantifies the change
in the extracted IRB error between the two ideally equivalent
constructions. Agreement within statistical uncertainty would
indicate that the comparison does not resolve an effect of the
explicit cancelling phase instructions under the tested
compilation and execution conditions.

In an ideal frame-update implementation, adjacent inverse virtual rotations restore the frame before the next physical pulse. This provides a natural equal-error baseline for PA-IRB. The paired comparison probes departures from this baseline in the complete compilation and execution workflow.

\section{Case Study: Transpiled Toffoli Gate}

We now apply PA-IRB to a transpiled Toffoli ($CCX$) gate as a representative
example of a compiled non-Clifford operation containing software-defined phase
updates. In this case study, the phase-stripped implementation is obtained by
removing the non-Clifford virtual $T$ and $T^\dagger$ phases from the frozen
compiled circuit, while the phase-dressed implementation retains each such
phase and immediately follows it by its inverse.

\subsection{Gate Construction and Layout}

Most quantum computing platforms do not support native three-qubit operations~\cite{barenco1995elementary,nielsen2002quantum}. The Toffoli (CCX) gate is one such example of a three-qubit gate that is not present in the native gate set of most quantum computing devices. Nevertheless, it is used in many algorithms and is supported by quantum libraries. The schematic representation of the Toffoli operation can be seen in Fig. \ref{scheme:tof}.
\begin{figure}
    \centering
\begin{quantikz}
\lstick{$q_0$} & \ctrl{1}  & \qw \\
\lstick{$q_1$} & \targ{}   & \qw \\
\lstick{$q_2$} & \ctrl{-1} & \qw
\end{quantikz}
    \caption{Toffoli configuration used in this work on a linear
    three-qubit topology $q_0-q_1-q_2$, with the outer qubits as controls
    and the middle qubit as the target.}
\label{scheme:tof}
\end{figure}

In order for the gate to be implemented on hardware that does not natively support it, it is often decomposed into a series of Clifford gates and non-Clifford $T/T^\dagger$ gates~\cite{barenco1995elementary,selinger2013tdepthone}. 

For the present experiment, the Toffoli gate is compiled on a local linear
three-qubit topology to the basis
$\{\mathrm{ID},R_Z,\sqrt{X},X,CZ\}$. The resulting frozen circuit is verified
to be exactly equivalent to the Toffoli unitary and contains 35 $R_Z$
rotations, 26 $\sqrt{X}$ gates, and 12 $CZ$ gates, with circuit depth 62.
Among the $R_Z$ rotations, seven are non-Clifford
$R_Z(\pm\pi/4)$ phases corresponding to virtual $T$ or $T^\dagger$
operations. These seven phase locations are the only instructions modified
when constructing the phase-stripped and phase-dressed PA-IRB circuits.

\subsection{PA-IRB Experimental Protocol}

For each physical qubit triple, we construct three matched circuit families:
reference RB, phase-stripped IRB, and phase-dressed IRB. For a given sequence
length and realization, the same random three-qubit Clifford sequence is used
to construct the corresponding reference, stripped, and dressed circuits.

The sequence lengths are
\[
m\in\{1,2,4,8,16,32,48\},
\]
with 15 random realizations for each sequence length and 2000 shots per
circuit. The random Clifford sequences and recovery operations are constructed
explicitly rather than using the built-in Qiskit Experiments IRB analysis.

To preserve the explicit virtual phase structure of the dressed circuits,
the final backend transpilation is performed with a fixed physical layout,
no additional routing, and optimization level zero. The dressed circuits are
checked after transpilation to ensure that the cancelling non-Clifford phase
pairs remain explicitly present and that no additional SWAP operations are
introduced.

\begin{figure*}[t]
    \centering
    \textbf{(a)}\quad
     \includegraphics[
        width=0.94\textwidth
    ]{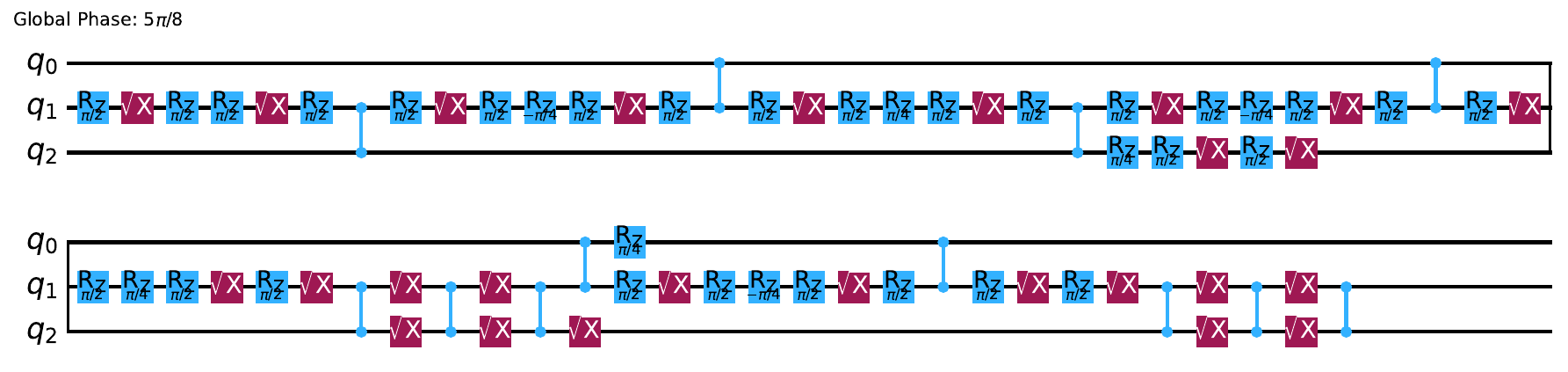}

    \vspace{0.4cm}

    \textbf{(b)}\quad
     \includegraphics[
        width=0.94\textwidth
    ]{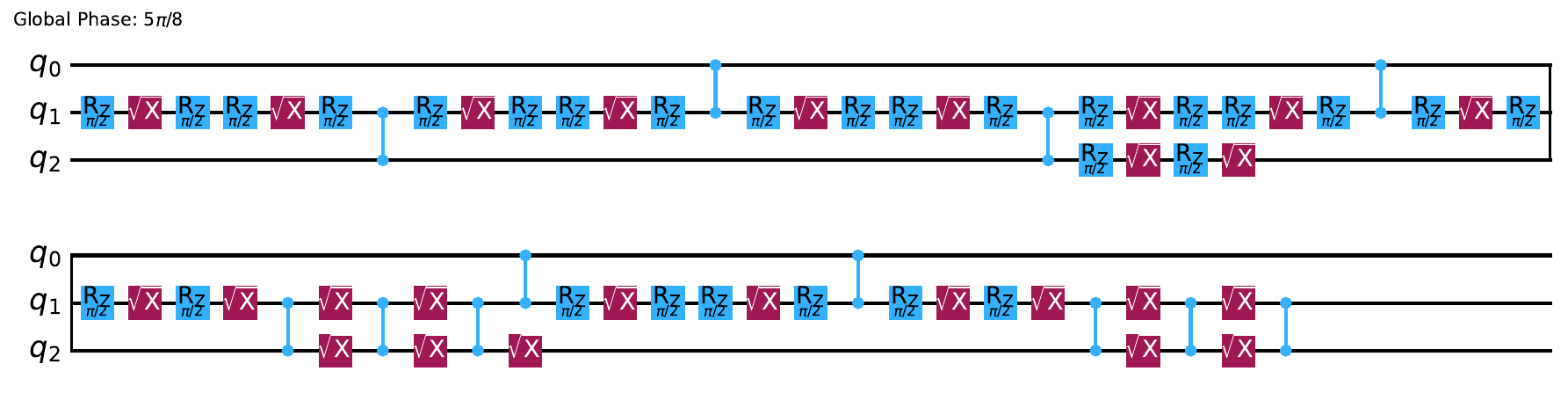}

    \vspace{0.4cm}

    \textbf{(c)}\quad
     \includegraphics[
        width=0.94\textwidth
    ]{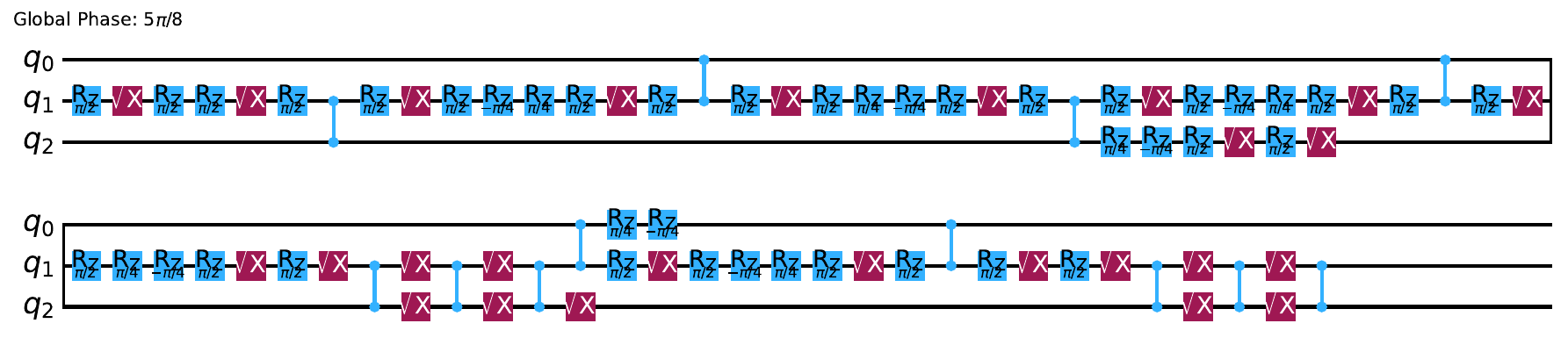}

    \caption{
    Compiled Toffoli construction used in the PA-IRB experiment.
    (a) Frozen compiled Toffoli gate containing seven non-Clifford
    virtual $R_Z(\pm\pi/4)$ rotations.
    (b) Phase-stripped construction $G_{\mathrm{ps}}$, obtained by
    removing these seven non-Clifford phase rotations while leaving the
    remaining instruction stream unchanged.
    (c) Phase-dressed construction $G_{\mathrm{pd}}$, in which each
    original non-Clifford phase rotation is retained and immediately
    followed by its inverse, forming an explicit $TT^\dagger$ or
    $T^\dagger T$ pair.
    The non-$R_Z$ physical gate skeleton is preserved between the
    phase-stripped and phase-dressed constructions.
    }
    \label{fig:toffoli_pairb_construction}
\end{figure*}

The preservation checks apply to the transpiled instruction stream. The resulting comparison characterizes the employed backend workflow, including any subsequent processing of the virtual phase instructions. The frozen compiled Toffoli and the corresponding phase-stripped and
phase-dressed constructions are shown in ~Fig.~\ref{fig:toffoli_pairb_construction}.

The experiment is performed on 215 qualified connected physical
qubit triples of \texttt{ibm\_kingston}. The population is fixed before data
collection and partitioned into six disjoint groups for parallel execution,
so that the tested qubit triples are not reselected between experimental runs.

\section{Results}

\subsection{Decay Fits and Error Estimates}

For each of the 215 physical qubit triples, the reference,
phase-stripped, and phase-dressed survival probabilities are averaged
over the 15 random realizations at each sequence length and fitted to the
decay model in Eq.~\ref{eq1}. The reference fit yields the decay parameter
$p$, while the phase-stripped and phase-dressed fits yield
$p_{\mathrm{ps}}$ and $p_{\mathrm{pd}}$, respectively. The corresponding
IRB error estimates $r_{\mathrm{ps}}$ and $r_{\mathrm{pd}}$ are then
obtained from Eq.~\ref{eq3}. 

\begin{figure*}[t]
    \centering

    \begin{minipage}[t]{0.32\textwidth}
        \centering
        \panelgraphic{\linewidth}{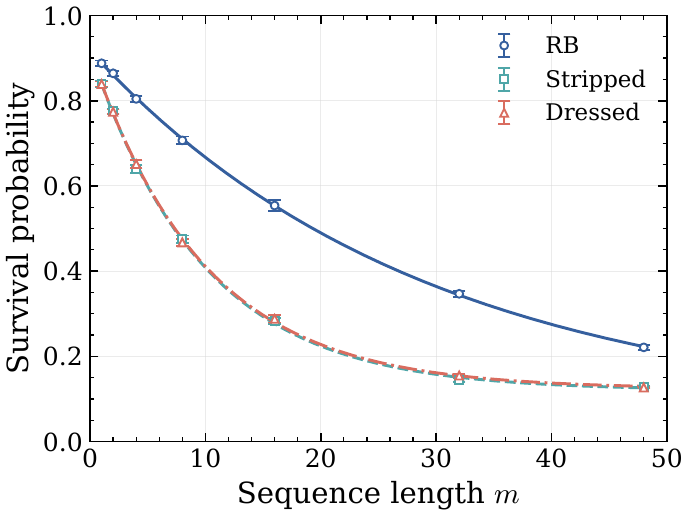}{(a)}
    \end{minipage}
    \hfill
    \begin{minipage}[t]{0.32\textwidth}
        \centering
        \panelgraphic{\linewidth}{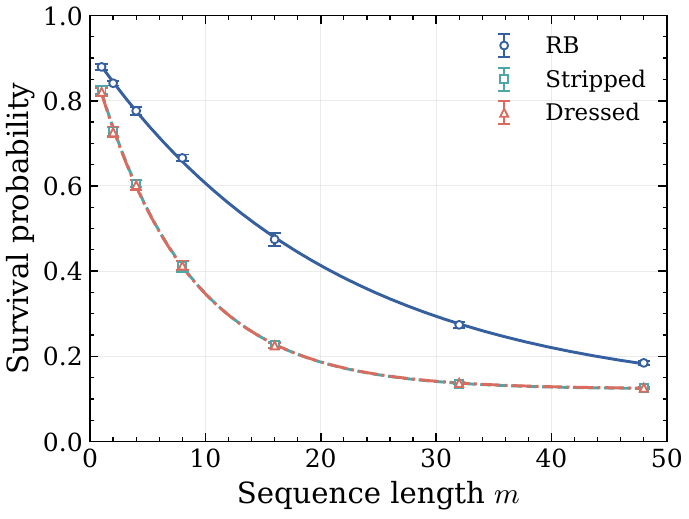}{(b)}
    \end{minipage}
    \hfill
    \begin{minipage}[t]{0.32\textwidth}
        \centering
        \panelgraphic{\linewidth}{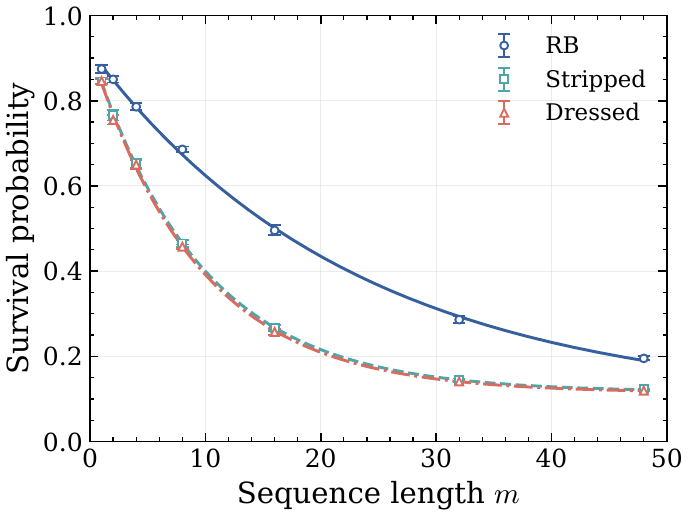}{(c)}
    \end{minipage}

    \caption{
    Representative PA-IRB decay curves for three physical qubit triples.
    Points show the mean survival probability over 15 random realizations
    at each sequence length, with error bars denoting the standard error
    of the mean across those realizations. Curves show fits to the decay
    model in Eq.~\ref{eq1} for reference RB, phase-stripped IRB, and
    phase-dressed IRB. The three examples illustrate fitted outcomes with
    (a) $\Delta r<0$, (b) closely matched phase-stripped and phase-dressed
    errors, and (c) $\Delta r>0$.
    }
    \label{fig:representative_decay}
\end{figure*}

Fits use unweighted nonlinear least squares with
$0\leq A,p,B\leq1$. Six initial values of the decay parameter,
$p_0\in\{0.50,0.70,0.85,0.93,0.97,0.99\}$, are tested, and
the solution with the smallest residual sum of squares is
retained. The error bars in Fig.~\ref{fig:representative_decay}
denote one standard error of the mean survival probability
across the 15 random realizations at each sequence length,
quantifying the sampling uncertainty of the averaged
experimental data.

For three qubits, Eq.~\eqref{eq3} gives $\Delta r=(7/8)(p_{\mathrm{ps}}-p_{\mathrm{pd}})/p$, so the paired difference uses a common reference decay parameter. The error bars displayed in Fig.~\ref{fig:representative_decay} describe the sampling uncertainty of the survival probabilities, while the population comparison below reports fitted point estimates.

The decay fits are well behaved for the large majority of the measured
triples. The median coefficients of determination are
$R^2=0.9993$, $0.9995$, and $0.9995$ for the reference,
phase-stripped, and phase-dressed fits, respectively. All reference fits
have $R^2>0.95$, while one pair
falls below this threshold. The threshold is used here only as a
fit-quality diagnostic; no triples are excluded from the primary
population analysis on this basis.

Representative decay curves are shown in
Fig.~\ref{fig:representative_decay}. The reference and interleaved data
show decreasing survival probability with increasing sequence length,
while the relative positions of the phase-stripped and phase-dressed
fitted decays vary between physical qubit triples. Cases occur in which
the phase-dressed construction yields the lower fitted error, cases in
which the phase-stripped construction yields the lower fitted error, and
cases in which the two fitted estimates are closely matched.

\subsection{Population Comparison of Phase-Stripped and Phase-Dressed Errors}

\begin{figure*}[t]
    \centering

    \begin{minipage}[t]{0.48\textwidth}
        \centering
        \panelgraphic{\linewidth}{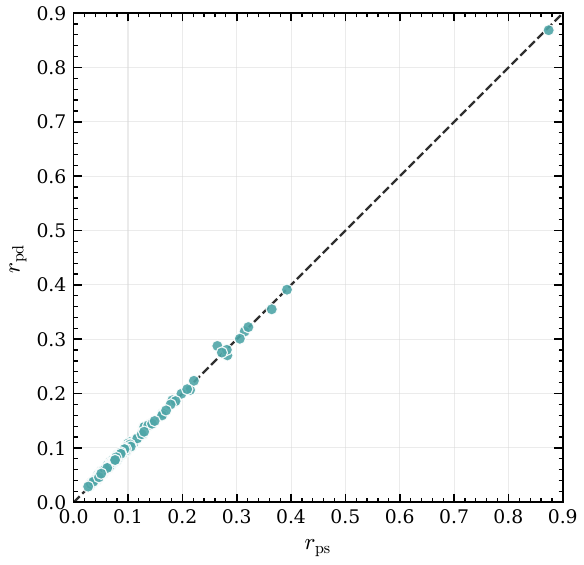}{(a)}
    \end{minipage}
    \hfill
    \begin{minipage}[t]{0.48\textwidth}
        \centering
        \panelgraphic{\linewidth}{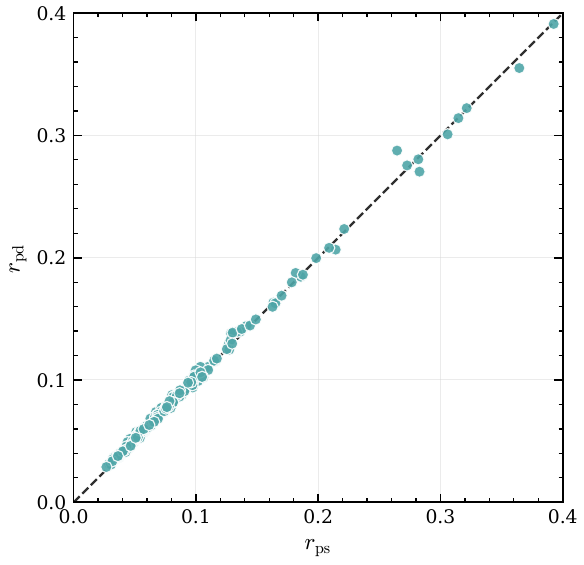}{(b)}
    \end{minipage}

    \caption{
    Population-scale comparison of the PA-IRB error estimates across
    215 physical qubit triples.
    (a) Full-range comparison of the fitted phase-dressed error
    $r_{\mathrm{pd}}$ with the fitted phase-stripped error
    $r_{\mathrm{ps}}$. The dashed diagonal marks
    $r_{\mathrm{pd}}=r_{\mathrm{ps}}$.
    (b) Expanded view of the low-error region, with both axes restricted
    to $0\leq r\leq0.4$.
    In panels (a) and (b), points below the diagonal correspond to
    $\Delta r<0$, while points above it correspond to $\Delta r>0$.
    }
    \label{fig:population_comparison}
\end{figure*}

Figure~\ref{fig:population_comparison} compares the fitted
$r_{\mathrm{ps}}$ and $r_{\mathrm{pd}}$ values across the full population
of 215 physical qubit triples. Each point represents one triple. The
diagonal $r_{\mathrm{pd}}=r_{\mathrm{ps}}$ corresponds to equal fitted
error for the phase-stripped and phase-dressed constructions. Points
below the diagonal have $\Delta r<0$, corresponding to a lower fitted
error for the phase-dressed construction, whereas points above the
diagonal have $\Delta r>0$, corresponding to a lower fitted error for
the phase-stripped construction. Because most fitted points occupy a
comparatively narrow low-error region, Fig.~\ref{fig:population_comparison}(b)
shows an expanded view of this part of the distribution.

Across the full population, 169 of the 215 fitted outcomes have
$\Delta r>0$, while 46 have $\Delta r<0$. The mean and median
fitted differences are $1.646\times10^{-3}$ and
$1.650\times10^{-3}$, respectively.
Thus, fitted outcomes occur on both sides of the equal-error diagonal,
while both the sample mean and median of $\Delta r$ are positive.

The observed point-estimate differences span
\[
    -1.249\times10^{-2}
    \leq
    \Delta r
    \leq
    2.312\times10^{-2}.
\]
The fitted differences span both signs, indicating that the observed stripped--dressed ordering varies across the tested triples. Here, this ordering is reported descriptively at the level of the fitted estimates.

All 215 preselected triples are retained in these population statistics.
In particular, the one phase-stripped fit and one phase-dressed fit with
$R^2<0.95$ are not removed retrospectively from the primary result. This
preserves the experimentally fixed population and avoids introducing a
fit-quality-based postselection criterion after data collection.

The measurements were acquired in six execution groups and span two
recorded backend calibration periods. Group~0 was executed before a
recorded calibration update, Groups~2--5 were executed afterward, and
the update occurred within the overall execution window of Group~1.
Each submitted job contains reference, phase-stripped, and phase-dressed
circuits, so the calibration transition is not associated exclusively
with one of the three circuit families. The population statistics
therefore characterize the complete acquisition window rather than a
single backend calibration snapshot.

Including all three circuit families in each job reduces their separation in acquisition conditions. Residual temporal variation can nevertheless contribute to the paired differences, and the population comparison is interpreted over the complete acquisition window.

\begin{figure*}[t]
    \centering

    \begin{minipage}[t]{0.48\textwidth}
        \centering
        \panelgraphic{\linewidth}{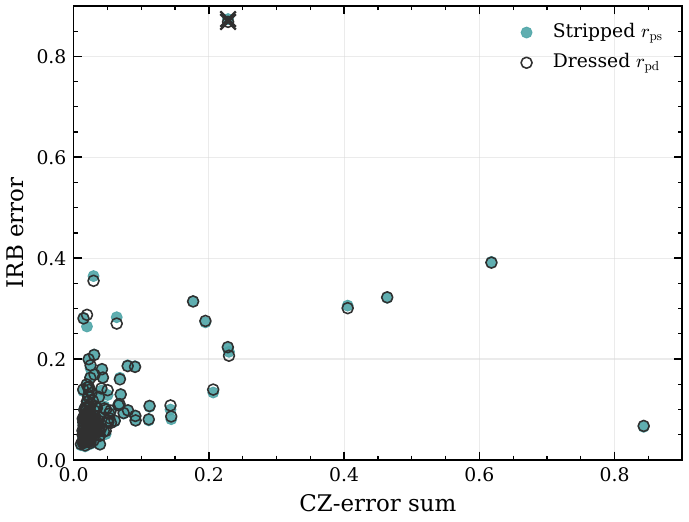}{(a)}
    \end{minipage}
    \hfill
    \begin{minipage}[t]{0.48\textwidth}
        \centering
        \panelgraphic{\linewidth}{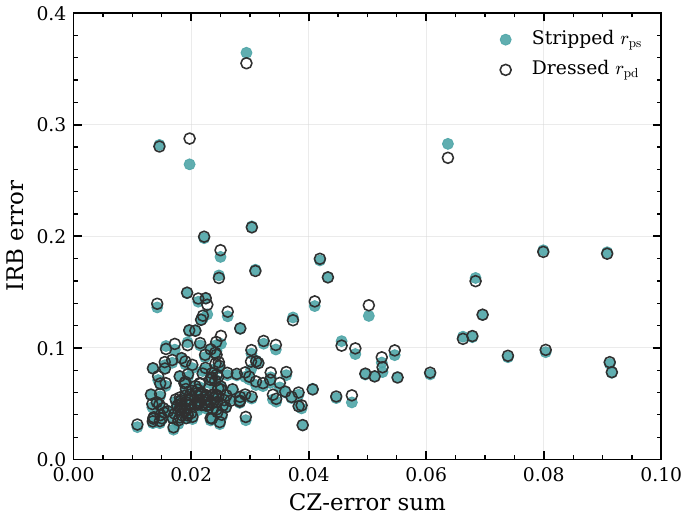}{(b)}
    \end{minipage}

    \vspace{0.3cm}

    \begin{minipage}[t]{0.48\textwidth}
        \centering
        \panelgraphic{\linewidth}{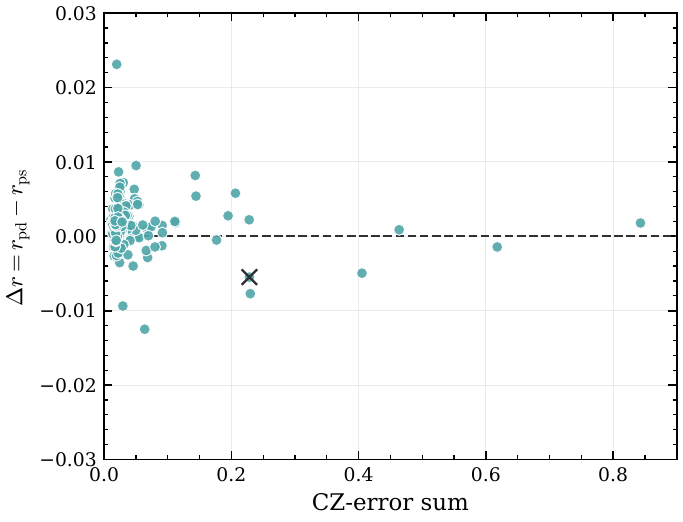}{(c)}
    \end{minipage}
    \hfill
    \begin{minipage}[t]{0.48\textwidth}
        \centering
        \panelgraphic{\linewidth}{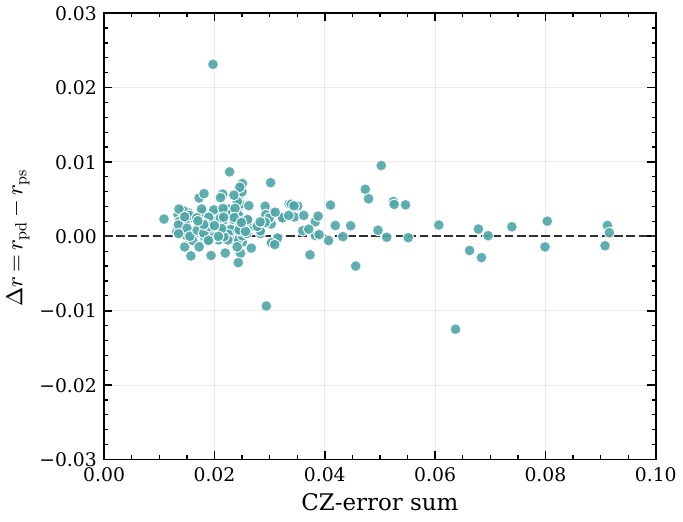}{(d)}
    \end{minipage}

    \caption{
    Comparison of the fitted PA-IRB error estimates with the
CZ-error sum.
    (a) Phase-stripped and phase-dressed error estimates across
    all 215 physical qubit triples.
    (b) Expanded view for CZ-error sums between $0$ and $0.10$,
containing 201 triples, with a reduced vertical range.
    (c) Signed difference
    $\Delta r=r_{\mathrm{pd}}-r_{\mathrm{ps}}$ across all
    215 triples.
    (d) Expanded view of the same 201 triples as in panel (b),
    retaining the vertical scale of panel (c).
    Dashed horizontal lines in panels (c) and (d) mark
    $\Delta r=0$.
    Crosses identify triples for which either surrogate fit has
    $R^2<0.95$; these observations are retained in the
    full-population analysis.
    }
    \label{fig:cz_error_comparison}
\end{figure*}

To relate the measured surrogate errors to the recorded device
calibration data, Fig.~\ref{fig:cz_error_comparison} compares both
error estimates and their signed difference with the CZ-error sum.
This quantity is calculated by summing the calibrated gate errors
over all 12 CZ instructions in the frozen circuit: four on
$(q_0,q_1)$ and eight on $(q_1,q_2)$, using the calibration values
associated with the triple's execution group. Both surrogates
preserve these multiplicities. The sum therefore serves as a
calibration-derived indicator, rather than an estimate of the
complete compiled-gate error.

Both surrogate error estimates show similar positive associations
with the CZ-error sum, with Pearson coefficients of approximately
$0.45$ and Spearman coefficients of approximately $0.47$.
For the signed difference $\Delta r$, the corresponding coefficients
are $-0.128$ and $-0.044$, respectively. These coefficients are
calculated over all 215 triples, including the fit-quality-flagged
triple. The CZ-error sum is therefore more strongly associated
with the overall surrogate error estimates than with their
fitted difference.

Taken together, the population data show a predominantly positive
$\Delta r$ at the level of the fitted point estimates, while retaining a
substantial set of triples for which $\Delta r<0$. PA-IRB therefore
characterizes a distribution of fitted stripped--dressed differences
across the tested qubit triples and acquisition window, rather than
a common fitted shift.

\section{Discussion}
\label{sec:discussion}

PA-IRB provides a paired diagnostic for compiled circuits containing software-defined phase operations. By comparing two constructions with the same ideal Clifford action and the same non-$R_Z$ gate skeleton, it makes the explicit virtual phase content an experimentally accessible feature of the benchmarking workflow. The Toffoli case study demonstrates this construction across 215 physical qubit triples using standard RB and IRB measurements.

The phase-stripped and phase-dressed estimates characterize Clifford surrogates of the original compiled operation. Their relation to the original non-Clifford gate error depends on the noise model, because phase alteration can modify the propagation of coherent errors. Under a model in which the background contribution is unchanged and phase-associated errors accumulate monotonically, the two estimates may motivate a lower--upper comparison. We retain this as a conditional interpretation rather than a bound established by the present measurements.

The population results are concentrated near equal surrogate errors, with a positive mean fitted difference and outcomes on both sides of zero. This pattern makes PA-IRB useful for identifying implementation-dependent variations that are obscured by a single device-wide average. The reported differences describe the tested compilation and acquisition conditions; attributing them to a specific physical mechanism lies beyond the paired comparison itself.

PA-IRB complements methods that directly characterize non-Clifford operations by offering a simple comparison within the Clifford IRB framework. Its use of matched random sequences and a shared reference provides a practical route to examining phase handling in compiled circuits without changing the non-$R_Z$ instruction skeleton.

The construction may also be applicable beyond IBM devices,
including trapped-ion~\cite{hughes2020benchmarking} and
neutral-atom~\cite{evered2023highfidelity,manetsch2025tweezer}
platforms employing software-defined phase operations, provided
that suitable Clifford surrogates and instruction-preservation
checks can be implemented. As quantum control stacks grow
deeper~\cite{vezvaee2025virtual}, benchmarking protocols must
become abstraction-aware.

\bibliographystyle{apsrev4-2}
\bibliography{irb-toffoli}

\end{document}